\documentclass[aps,prl,reprint,superscriptaddress]{revtex4-1}
\usepackage{graphicx,graphics,amsfonts,amsmath,amssymb}

\usepackage{color}
\usepackage{threeparttable}

\begin{document}

% Use the \preprint command to place your local institutional report
% number in the upper righthand corner of the title page in preprint mode.
% Multiple \preprint commands are allowed.
% Use the 'preprintnumbers' class option to override journal defaults
% to display numbers if necessary
%\preprint{}

%Title of paper
\title{Quantum degenerate Bose-Fermi mixture of chemically different atomic species with widely tunable interactions}

% repeat the \author .. \affiliation  etc. as needed
% \email, \thanks, \homepage, \altaffiliation all apply to the current
% author. Explanatory text should go in the []'s, actual e-mail
% address or url should go in the {}'s for \email and \homepage.
% Please use the appropriate macro foreach each type of information

% \affiliation command applies to all authors since the last
% \affiliation command. The \affiliation command should follow the
% other information
% \affiliation can be followed by \email, \homepage, \thanks as well.
\author{Jee Woo Park}
\author{Cheng-Hsun Wu}
\author{Ibon Santiago}
\affiliation{MIT-Harvard Center for Ultracold Atoms, and Research Laboratory of Electronics, Massachusetts Institute of Technology,
Cambridge, Massachusetts 02139, USA }
\affiliation{Department of Physics, Massachusetts Institute of Technology,
Cambridge, Massachusetts 02139, USA}
\author{Tobias G. Tiecke}
\affiliation{MIT-Harvard Center for Ultracold Atoms, and Research Laboratory of Electronics, Massachusetts Institute of Technology,
Cambridge, Massachusetts 02139, USA }
\affiliation{Department of Physics, Harvard University, Cambridge, Massachusetts 02138, USA}
\author{Sebastian Will}
\author{Peyman Ahmadi}
\author{Martin W. Zwierlein}
\affiliation{MIT-Harvard Center for Ultracold Atoms, and Research Laboratory of Electronics, Massachusetts Institute of Technology,
Cambridge, Massachusetts 02139, USA }
\affiliation{Department of Physics, Massachusetts Institute of Technology,
Cambridge, Massachusetts 02139, USA}
%PETER: Physics Department->Department of Physics

%\email[]{Your e-mail address}
%\homepage[]{Your web page}
%\thanks{}
%\altaffiliation{}

%Collaboration name if desired (requires use of superscriptaddress
%option in \documentclass). \noaffiliation is required (may also be
%used with the \author command).
%\collaboration can be followed by \email, \homepage, \thanks as well.
%\collaboration{}
%\noaffiliation

\date{\today}

\begin{abstract}
We have created a quantum degenerate Bose-Fermi mixture of $^{23}$Na and $^{40}$K with widely tunable interactions via broad interspecies Feshbach resonances. Over 30 Feshbach resonances between $^{23}$Na and $^{40}$K were identified, including $p$-wave multiplet resonances. The large and negative triplet background scattering length between $^{23}$Na and $^{40}$K causes a sharp enhancement of the fermion density in the presence of a Bose condensate. As explained via the asymptotic bound-state model, this strong background scattering leads to wide Feshbach resonances observed at low magnetic fields. Our work opens up the prospect to create chemically stable, fermionic ground state molecules of $^{23}$Na--$^{40}$K where strong, long-range dipolar interactions would set the dominant energy scale.
\end{abstract}
% insert suggested PACS numbers in braces on next line
% insert suggested keywords - APS authors don't need to do this
%\keywords{}

\maketitle

Ultracold quantum gases realize paradigms of condensed matter physics in pristine fashion, such as the superfluid to Mott insulator transition~\cite{bloc08review}, the BEC-BCS crossover in fermionic superfluids~\cite{ingu08varenna,kett08rivista} and the Berezinskii-Kosterlitz-Thouless transition in two-dimensional Bose gases~\cite{hadz11rivista}. A plethora of novel many-body systems may become accessible through the advent of quantum mixtures of different atomic species. In particular, Bose-Fermi mixtures with widely tunable interactions should reveal boson mediated interactions between fermions and possibly boson induced $p$-wave superfluidity~\cite{Bijlsma2000,Heisel2000}. The fate of impurities in a Fermi sea~\cite{schi09polaron} or a Bose condensate~\cite{Wu2011, Fratini2010, Yu2011} can be studied, and new quantum phases of matter are predicted in optical lattices~\cite{Lew2004}.
Furthermore, the creation of fermionic ground state molecules starting from a degenerate Bose-Fermi mixture opens up a whole new avenue of research, as this results in a Fermi gas with long-range, anisotropic dipole-dipole interactions~\cite{ni09polar}.
Since the first degenerate Bose-Fermi mixture of different atomic species, $^{23}$Na and $^6$Li~\cite{Hadzibabic2002}, a variety of such systems has been realized~\cite{Roati2002,gold04,ospe05,aubi05bosefermi,gunt06bosefermi,best09bosefermi,Silber2005, Taglieber2008,hara11mixture, Wu2011,hans11yttli}. However, so far only one mixture, $^{87}$Rb-$^{40}$K, has allowed tunability of interspecies interactions with relative ease by means of a moderately wide ($\Delta B \approx 3\,\rm G$) Feshbach resonance~\cite{ferl06krb}, and only in this case fermionic Feshbach molecules have successfully been produced~\cite{ospe06hetero,zirb08hetero}.

\begin{figure}
\begin{center}
\includegraphics[width=3.2in]{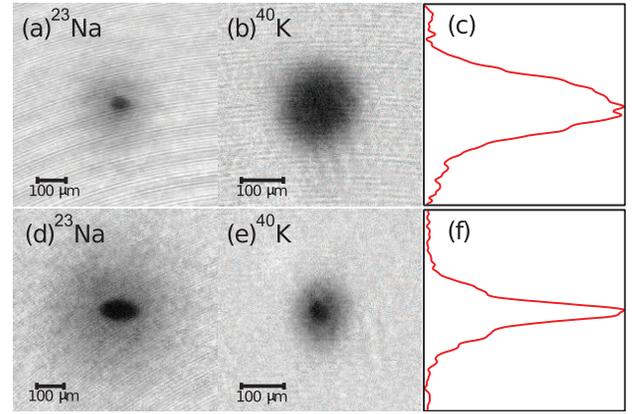}
\end{center}
\setlength\abovecaptionskip{0pt}
\caption {(Color online) Simultaneous quantum degeneracy of $^{23}$Na and $^{40}$K atoms. (a)-(b) and (d)-(e) are pairs of time of flight (TOF) absorption images of a $^{23}$Na BEC and a $^{40}$K Fermi cloud with different atom number balance.
A strong attractive interaction between the two species is observed in (e) as a sharp increase of the central density in the fermionic cloud in the presence of a Bose condensate. (c) and (f) are the center-sliced column density of the fermionic clouds of (b) and (e), respectively. TOF was (a) 11 ms, (b) 7 ms, (d) 17 ms, and (e) 5 ms. Atom numbers are (a) $2.4\cdot 10^5$, (b) $2 \cdot 10^5$, (d) $8.2 \cdot 10^5$, and (e) $6.7 \cdot 10^4$.}
\label{degeneracy}
\end{figure}

In this article, we report on the experimental realization of a new Bose-Fermi mixture of $^{23}$Na and $^{40}$K and the observation of over 30 $s$- and $p$-wave Feshbach resonances at low magnetic fields. We demonstrate that $^{23}$Na is an efficient coolant for sympathetic cooling of $^{40}$K. A pattern of wide $s$-wave resonances exists for most of the energetically stable hyperfine combinations, the widest being located at 138 G with a width of about 30 G in the $^{23}$Na$|F=1,m_F=1\rangle$+$^{40}$K$|F=9/2,m_F=-5/2\rangle$ hyperfine configuration. We also observe $p$-wave multiplet resonances that are resolved thanks to their location at low magnetic fields.

In the singlet rovibrational ground state, the NaK molecule is known to have a large permanent electric dipole moment of 2.72(6) D~\cite{worm81nak,gerd11nak}, five times larger than that of KRb~\cite{ni09polar}, and is predicted to be chemically stable against atom-atom exchange reactions~\cite{Hutson2011}, in contrast to KRb~\cite{ospe2010}. An ultracold gas of fermionic ground state molecules of NaK will thus be an ideal system for the study of Fermi gases with strong, long-range dipolar interactions. Indeed, the interaction energy here can be expected to be on the order of the Fermi energy.

The experimental setup is based on the apparatus presented in Ref.~\cite{Wu2011}, which employs two independent Zeeman slowers capable of simultaneously loading sodium and potassium atoms directly into a UHV chamber. The potassium slower selectively decelerates the $^{40}$K isotope, loading $10^7$ atoms into a magneto-optical trap (MOT) in 10 s. For $^{23}$Na, we use a dark spot MOT~\cite{WK93}, which allows us to load approximately $10^9$ atoms in 2 s.

Multi-species experiments can suffer from atom losses due to light-assisted collisions in the MOT and spin-changing collisions in the magnetic trap. In order to minimize such losses, we developed a shelving technique where $^{23}$Na is first loaded into the MOT, optically pumped to the $|2,2\rangle$ stretched state, and captured in the magnetic trap. Next, the trap gradient is reduced to 7.7 G/cm to only support the stretched state of $^{23}$Na against gravity. With this gradient left on, and the $^{23}$Na thus ``shelved in the dark'' in a purely magnetic trap, finally the MOT and slower beams for $^{40}$K are switched on to load the $^{40}$K MOT. This scheme guarantees that only $^{23}$Na atoms in the $|2,2\rangle$ state are present in the magnetic trap before loading $^{40}$K, and it also potentially reduces light-assisted collisions that would be encountered in a double-species MOT.

Once both species are loaded into the optically plugged magnetic trap~\cite{Wu2011}, the mixture is cooled for 7 s by rf-induced evaporation of $^{23}$Na, where thermally excited $^{23}$Na atoms in the $|2,2\rangle$ state are removed from the trap by coupling to the high field seeking state of $|1,1\rangle$. We decompress the initial magnetic gradient of 220 G/cm to 27.5 G/cm at the end of evaporation to reduce three-body losses. The 5 $\mu$K cold mixture is then loaded into a crossed optical dipole trap (laser wavelength $1064$ nm, maximum power 4.7 and 17 W, waist 73 and 135 $\mu$m).

At this stage, the $1/e$ lifetime of the mixture, with $^{23}$Na and $^{40}$K still in their stretched states, is about $\tau=250$ ms, already signaling a strong attractive interaction increasing three-body losses and spin-changing dipolar losses. We thus immediately transfer $^{23}$Na atoms into their hyperfine ground state $|1,1\rangle$ via a Landau-Zener sweep, and remove any remaining $|2,2\rangle$ atoms via a resonant light pulse. In the $^{23}$Na$|1,1\rangle$+$^{40}$K$|9/2,9/2\rangle$ state, the mixture now lives for $\tau=20$ s. The gas is further evaporatively cooled in this spin mixture for 2 s by reducing the intensity of the dipole trap beams.

At the end of evaporation, a degenerate Fermi gas of $^{40}$K with $2 \times 10^5$ atoms and $T/T_F=0.6$ coexists with a Bose-Einstein condensate of $^{23}$Na. Two sets of absorption images of the mixture for different values of $^{23}$Na and $^{40}$K atom numbers are shown in Fig.~\ref{degeneracy}, where atom numbers are varied by changing the MOT loading times of the two species. The strong attractive interaction between $^{23}$Na and $^{40}$K in the $^{23}$Na$|1,1\rangle$+$^{40}$K$|9/2,9/2\rangle$ state is apparent. As the condensate grows, the fermionic cloud acquires a bimodal density distribution as it experiences the strong mean-field potential of the bosons~\cite{ospe06hetero}--see Fig. \ref{degeneracy}(e).

\begin{figure}
\begin{center}
\includegraphics[width=3.2in]{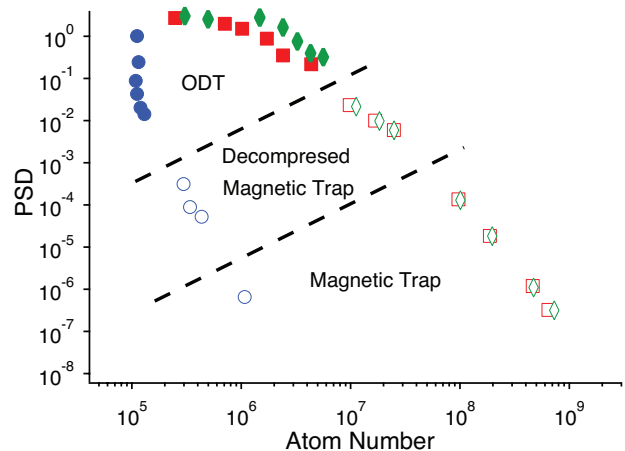}
\end{center}
\setlength\abovecaptionskip{-3pt}
\caption {(Color online) Evolution of the phase space density (PSD) with atom number ($N$). Blue circles: $^{40}$K; red squares: $^{23}$Na with $^{40}$K;
green diamonds: $^{23}$Na alone. Open and solid symbols represent the PSD in the magnetic trap and optical trap, respectively.} \label{PSD}
\end{figure}

The evolution of the phase space densities (PSDs) and atom numbers $N$ of both species during evaporation is shown in Fig.~\ref{PSD}. Temperature is determined by fitting a thermal profile to the wings of the $^{23}$Na cloud. The cooling efficiency $\Gamma = - d \ln(\text{PSD})/d\ln(N)$ for sodium in the magnetic trap is $\Gamma_{\text{Na}}=2.7$, a rather high value thanks to the steep confinement in the plugged trap. This efficiency is not affected by the presence of the relatively small admixture of $^{40}$K. Sympathetic cooling is less efficient than in other mixtures~\cite{hadz03big_fermi,Wu2011} as $^{40}$K is seen to be lost due to three-body collisions in the magnetic trap. We find $\Gamma_{\text{K}}=4.6$ for $^{40}$K. In the crossed optical dipole trap, with sodium in $|1,1\rangle$, the sodium cooling becomes less efficient due to the weaker confinement, $\Gamma_\text{Na}=0.9$, but the $^{40}$K number is essentially conserved in this mixture so that sympathetic cooling is highly efficient, with $\Gamma_\text{K}=15.3$. The lowest $T/T_F$ achieved for $^{40}$K after evaporating all of $^{23}$Na is $T/T_F$ = 0.35 with $3 \times 10^5$ atoms.

\begin{table}
\caption{\label{tab:table1} Data summary on the Feshbach resonances between $^{23}$Na in $|1,1\rangle$ and $^{40}$K in $|9/2,m_F\rangle$. The positions and widths of the resonances, $B^{\rm expt}_0$ and $\Delta B_{\rm expt}$, are determined by phenomenological Gaussian fits ($\propto e^{-(B-B_{0}^{\rm expt})^{2}/ \Delta B_{\rm expt}^2}$) to the observed loss features. For $p$-wave resonances, we report the  positions and widths of the resolved features, i.e., doublets and multiplets. $B_0^{th}$  and $\Delta B_{th}$ give the theoretical positions and widths from the ABM model.}
\begin{threeparttable}
\begin{ruledtabular}
\setlength{\tabcolsep}{1pt}
\begin{tabular}{cccccc}
$^{23}$Na$|1,1\rangle$+ & $B^{\rm expt}_0$&$\Delta B_{\rm expt}$& $B^{\rm th}_0$&$\Delta B_{\rm th}$&Res.\\
$^{40}$K$|9/2,m_F\rangle$&(G)&(G)&(G)&(G) &  type \\
-9/2& 6.35, 6.41, 6.47, 6.68 & 0.02& 7.2 &&$p$\\
	& 19.12, 19.18, 19.27 &0.02 & 18.3 &&$p$\\
 	& 78.3 &1.1& 75.5 & 1.1 & $s$\\
	& 88.2 &4.3& 84.5 & 5.4 & $s$\\

-7/2 & 7.32, 7.54 &0.2, 0.03 & 8.7 &&$p$\\
	 & 23.19, 23.29 &0.05, 0.05& 22.1 &&$p$\\
	 & 81.6  &0.2 & 82.1 & 0.04 & $s$\\
	 & 89.8  &1.1 & 87.3 & 0.6 & $s$\\
	 & 108.6 &6.6 & 105.7 & 13.1 & $s$\\

-5/2 & 9.23, 9.60 &0.14, 0.11& 11.0 &&$p$\\
	 & 29.19, 29.45, 29.52 & 0.04& 27.8 &&$p$\\
	 & 96.5  &0.5 & 97.2 & 0.04 & $s$\\
	 & 106.9 &1.8 & 103.8 & 0.45 & $s$\\
	 & 148 (138\tnote{*} ) & 37 (30\tnote{*} ) & 137.1 & 26 & $s$\\

-3/2 & 12.51, 12.68 &0.16, 0.06& 14.8 &&$p$\\
	 & 39.39, 39.86 &0.15, 0.14& 37.2 &&$p$\\
	 & 116.9  &0.5 & 118.3 & 0.07 & $s$\\
	 & 129.5 &4.6 & 127.2 & 0.39 & $s$\\
	 & 175 &20 & 187.8 & 50.5 & $s$\\

\end{tabular}
\begin{tablenotes}
\item[*] The resonance position and width have been refined by measuring the molecular binding energies via rf spectroscopy~\cite{wu12mol}.
\end{tablenotes}
\end{ruledtabular}
\end{threeparttable}
\end{table}

With a new degenerate Bose-Fermi mixture at our disposal, the natural next step is to search for interspecies Feshbach resonances between $^{23}$Na and $^{40}$K. There have been theoretical indications for resonances below 100 G~\cite{gerd08nak}. In addition, a large and negative triplet scattering length was predicted in Ref.~\cite{vent01}, a value that has recently been refined to $a_t = -575^{+191}_{-532}$~\cite{gerd08nak}, indicating that the triplet potential has an almost bound, virtual state right above threshold. A large background scattering length is often a catalyst for wide Feshbach resonances~\cite{marc04res}, caused by strong coupling of molecular states to the almost resonant open channel. A famous example is the 300 G wide Feshbach resonance in $^6$Li~\cite{diec02fesh}.

\begin{figure}
\begin{center}
\includegraphics[width=3.3in]{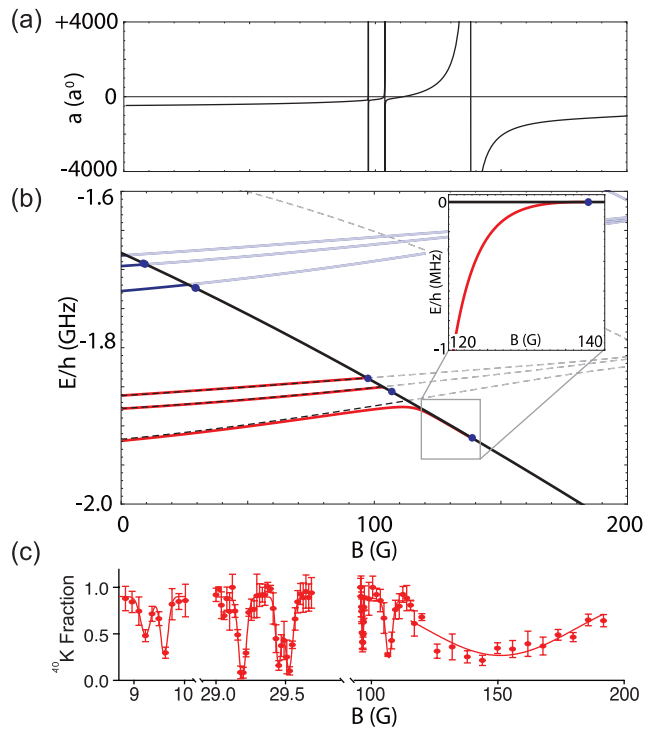}
\end{center}
\setlength\abovecaptionskip{-2pt}
\caption {(Color online) Feshbach resonances in $^{23}$Na-$^{40}$K, here for $^{23}\text{Na}|1,1\rangle+  ^{40}\text{K}|9/2,-5/2\rangle$ collisions. (a) Scattering length from the ABM model. (b) Open channel threshold energy (black solid line), uncoupled ($s$-wave: dashed lines, $p$-wave: light blue lines) and coupled molecular states ($s$-wave: red solid lines, $p$-wave: blue solid lines). The blue dots denote experimentally measured resonances. The inset shows the energy, relative to threshold, of the molecular state at the wide $s$-wave resonance at 138 G~\cite{wu12mol}. (c) Experimental loss spectra of $^{40}$K in the presence of $^{23}$Na. Three $s$-wave resonances and two $p$-wave manifolds are found,  the latter resolved in one doublet and one triplet.} \label{feshbach}
\setlength\belowcaptionskip{-1cm}
\end{figure}

We performed Feshbach loss spectroscopy, mapping out atom losses of both species as a function of magnetic field. Over 30 Feshbach resonances were observed in four different spin state combinations of $^{23}$Na$|1,1\rangle$+$^{40}$K$|9/2,m_{F}\rangle$, from the ground spin state $m_{F}=-9/2$ up to $m_{F}=-3/2$. Spin states of $^{40}$K are prepared starting from $m_F = +9/2$ by a single Landau-Zener sweep through the intermediate $m_{F}$ states at 15 G. The experimentally observed resonance positions and widths are reported in Table~\ref{tab:table1}. Many wide $s$-wave Feshbach resonances at low magnetic fields are identified, the widest one at 138 G for collisions between $^{23}\text{Na}|1,1\rangle+  ^{40}\text{K}|9/2,-5/2\rangle$, with a width of $30$ G--see Fig.~\ref{feshbach}. This is an order of magnitude wider than any other resonances found so far in a Bose-Fermi mixture of chemically different atomic species.

\begin{figure}
\begin{center}
\includegraphics[width=3.4in]{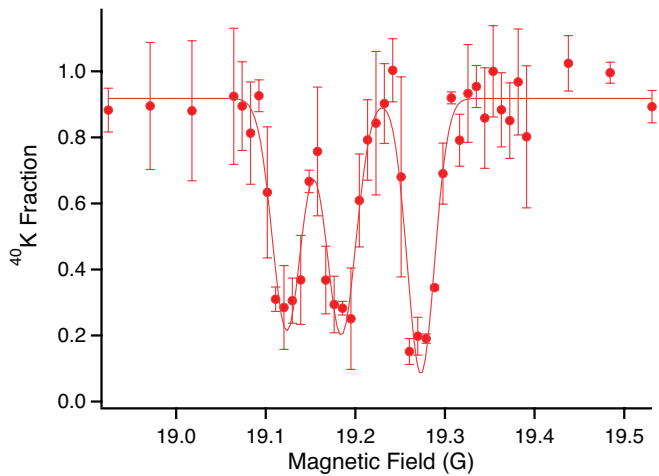}
\end{center}
\setlength\abovecaptionskip{-3pt}
\caption {(Color online) Triplet structure of the $p$-wave resonance at 19.1 G for the $^{23}$Na$|1,1\rangle$+$^{40}$K$|9/2,-9/2\rangle$ spin configuration. A phenomenological triple Gaussian fit is applied as a guide to the eyes.} \label{pwave}
\end{figure}

$p$-wave Feshbach resonances are known to split into a doublet structure due to different projections of the orbital angular momentum onto the magnetic field axis \cite{tick04pwave}.
In the NaK system, however, we observe triplet features for many $p$-wave resonances--see Fig~\ref{pwave}. These originate from the magnetic dipole-dipole interaction of constituent atoms, which couple molecular states with different total internal spin. The diagonal terms of the magnetic dipole-dipole interaction induce an energy shift that differs for the $m_l=0$ and $|m_l|=1$ quantum numbers, giving rise to the well known doublet structures. The off-diagonal terms in the dipole-dipole interaction couple different values of $m_l$ while conserving the total angular momentum $m_l + M_F$, where $M_F=m_{F,\text{K}}+m_{F,\text{Na}}$. These terms are in most mixtures negligible since molecular states with different values of $M_F$ have to be nearly degenerate to result in a significant energy shift. However, due to the low-field nature of the NaK $p$-wave resonances, multiple molecular states are nearly degenerate with the open channel spin state, allowing for the triplet structure to be resolved.

The assignment of $s$- and $p$-wave characters of the resonances follows from a simple, but powerful model of the molecular states involved. The singlet and triplet potentials of the interatomic potential allow for a variety of bound states. From the known scattering length~\cite{gerd08nak} and the van-der-Waals coefficient $C_6$~\cite{dere01c6}, the weakest $s$-wave bound states are expected at about $E_s^s =-150(10)\,\rm MHz$ and $E_t^s = -1625(50)\,\rm MHz$ for the singlet and triplet $s$-wave potentials, where the errors reflect uncertainties of the scattering lengths from Ref.~\cite{gerd08nak}. The $p$-wave bound states follow from the $s$-wave bound states as in Ref.~\cite{tiecke2010} and are slightly adjusted to fit the observed resonance positions.
As described in Refs.~\cite{moer95res,stan04}, as a first guess for locations of Feshbach resonances one can neglect the part of the hyperfine interaction that couples singlet and triplet bound states. This already provides the pattern of Feshbach resonances as positions where these (uncoupled) molecular states cross the atomic (open-channel) threshold. The analysis shows that the observed resonances are caused by the triplet bound states. Next, we use the asymptotic bound state model (ABM) to include the singlet-triplet coupling of molecular states~\cite{will07fermifermi}. To couple the molecular states to the open channel, we follow Marcelis \emph{et al.}~\cite{marc04res} and only include the effect of the virtual state causing the large and negative triplet scattering length. The spin part of the coupling matrix element is obtained from the ABM Hamiltonian and the spatial part, i.e., the wavefunction overlap between the respective bound state and the virtual state, is taken as one free fit parameter. For the background scattering length of the low-field resonances the effect of broad resonances is included. The virtual state causes strong coupling of several $s$-wave molecular states to the open channel, leading to wide, open-channel dominated resonances as known from the case of $^6$Li. The theoretical values obtained with this model are shown in Table~\ref{tab:table1}. An exceptionally broad resonance for $m_F = -3/2$ is predicted to be even wider and to be shifted further, possibly hinting at a shift between the loss maximum and the actual Feshbach resonance position. Our approach leads to a refined triplet bound state energy of $E_t^s = -1654(3)\,\rm MHz$ and $E_t^p = -1478(7)\,\rm MHz$, and using the long range potential from Ref.~\cite{gerd08nak}, we find a refined value of the triplet scattering length of $a_t = -830(70)\, a_0$. The errors correspond to one standard deviation of a least-squares fit to the eight narrowest $s$-wave resonances that are least sensitive to the coupling to the scattering states and hence the error induced by the ABM is expected to be small~\cite{will07fermifermi}.

In conclusion, we have produced a degenerate Bose-Fermi mixture of $^{23}$Na and $^{40}$K, and identified over 30 $s$- and $p$-wave interspecies Feshbach resonances, including several exceptionally broad resonances. Remarkably, many $p$-wave Feshbach resonances are observed to be triplets or even multiplets. Our strongly interacting $^{23}$Na-$^{40}$K mixture near these Feshbach resonances should allow the study of Bose or Fermi polarons~\cite{schi09polaron}, of boson mediated interactions between fermions, and possibly of novel states of matter in optical lattices. The formation of fermionic Feshbach molecules is within reach. In the rovibrational ground state, NaK molecules possess a large induced electric dipole moment and are stable against exchange reactions. One can thus hope to create a Fermi gas of polar molecules with strong dipole-dipole interactions that dominate the many-body physics of the gas, rather than being a small perturbative effect.

We thank Eberhard Tiemann for performing a coupled-channel calculation based on our results which stimulated us to further investigate the $p$-wave multiplet structures. We also thank Tout Wang for stimulating discussions. This work was supported by the NSF, AFOSR-MURI and -PECASE, ARO-MURI, ONR YIP, DARPA YFA, a grant from the Army Research Office with funding from the DARPA OLE program and the David and Lucille Packard Foundation.

\bibliographystyle{apsrev}
\bibliography{referencesmerged}

\end{document}